\documentclass[twocolumn,aps,prl,10pt,amsmath,amssymb,nofootinbib,showpacs,superscriptaddress,floatfix]{revtex4-1}

\DeclareFontFamily{U}{rcjhbltx}{}
\DeclareFontShape{U}{rcjhbltx}{m}{n}{<->rcjhbltx}{}
\DeclareSymbolFont{hebrewletters}{U}{rcjhbltx}{m}{n}

\usepackage{graphicx}
\usepackage{color}
\usepackage[usenames,dvipsnames]{xcolor}
\usepackage[colorlinks=true,linkcolor=Red,citecolor=Green,linktoc=page]{hyperref}
\usepackage{multirow}
\usepackage{float}
\usepackage{flushend}
\usepackage{balance}
\usepackage[varg]{txfonts}
\usepackage{ulem}
\usepackage{fancyhdr}

\DeclareMathSymbol{\lamed}{\mathord}{hebrewletters}{108}

\begin{document}
\title{Complexity and precision in quantum associative retrieval}



\author{C.\,A.\,Trugenberger}

\affiliation{SwissScientific Technologies SA, rue du Rhone 59, CH-1204 Geneva, Switzerland.}
\affiliation{Division of Science, New York University Abu Dhabi, Abu Dhabi, United Arab Emirates.}
\email{E-mail: ca.trugenberger@bluewin.ch}

\begin{abstract}
We show that efficient and high-precision quantum associative retrieval is possible with an exponential complexity advantage with respect to the classical brute force approach and with a polynomial capacity advantage with respect to classical neural network implementations for any number of patterns scaling polynomially in the number of qubits. Both classical neural and quantum associative memories store the patterns in a Hamiltonian; in the classical case the dynamics is irreversible relaxation toward attractors, in the quantum case it is unitary transformations generating an effective thermal distribution, followed by a measurement. 

\end{abstract}
\maketitle


Associative pattern recognition is the task of retrieving the memory item stored in a device which is closest, with respect to a predefined distance, to a given input, i.e., given a set of $p$ binary strings of length $n$ and an input string, the task is to retrieve one of the stored patterns with minimal distance from the input, with high probability and reduced computational complexity.
The brute force approach based on RAM memories is very inefficient for a large number of patterns $p$, since the input has to be compared sequentially to each one of the stored memories, leading to a complexity $O(pn)$.  Associative retrieval, however, is one of the defining capabilities of the human brain. Its implementation on neural networks \cite{hopfield}, which mitigates the brute force complexity, was awarded the 2024 Nobel prize. 

Unfortunately, even this approach is plagued by a problem; the number of retrievable patterns on $n$ bits has a limit $O(n)$ (for a review see \cite{muller}). For larger number of memories the phenomenon of cross-talk leads to a phase transition generating a completely confused state of the network, in which all retrieval capabilities are lost. 

Around the turn of the century, two authors \cite{ventura, cat1, cat2, cat3} considered the quantum version of associative retrieval, to establish if both the brute force complexity and the cross-talk limitations of neural network implementations could be avoided and quantum memories allowing efficient associative retrieval for at least a polynomial number of patterns could be constructed. Closely related to this topic, the quantum classification task was considered in \cite{sasaki} and the quantum identification of regularities in \cite{schutzhold}. All these works can be considered the birth of quantum machine learning. The quantum associative retrieval problem was reconsidered, in different variants, more than a decade later \cite{petruccione, lloyd2, sousa, khan}. More recently, the whole field of machine learning is being extended to the quantum domain \cite{biamonte} (for a review see \cite{briegel}). 

The na\"ive direct quantization of the Hopfield model, mapping neurons to qubits, does not improve the $O(n)$ capacity limit \cite{dt}. The quantum advantage emerges when 
patterns are stored in quantum states. An algorithm to store a given set of binary patterns in a uniform quantum superposition was first proposed in \cite{ventura}. A simpler version, which can be implemented in a quantum circuit (for a review see \cite{nielsen}) was then suggested in \cite{cat1}. A fundamental limitation on storing patterns directly in quantum states, however is imposed by the no-cloning theorem \cite{zurek}: once the memory state is measured for retrieval, it is irreversibly destroyed and cannot be accessed again. The superposition of stored patterns, therefore, should be viewed only as a transient initial state. 

The real quantum memory resides in the hardware that implements the unitary transformation which prepares this initial state from a fixed reference input. A sequential quantum circuit with $O(p)$ gates, however is not suitable to attain a complexity much smaller than the classical brute force lookup. A genuine associative memory must store information in a static physical structure, not recreate it from scratch each time it is queried. This is exactly how the classical Hopfield network operates: the patterns are embedded in the synaptic weight matrix, a spatially fixed object. Note that storing patterns in a quantum superposition, or in the hardware implementation that creates it in a single pass, does not change the functional definition of ``associative memory", retrieval is still based on content similarity rather than storage address identification.

Fortunately, such a ``single-pass" hardware implementation of unitary transformations exists, in the form of photonic interferometers \cite{reck}. By ``single-pass" we mean that the transformation is implemented by a fixed physical device in one traversal, without a sequence of gates whose number scales with the number of stored patterns. In an interferometric quantum memory, the patterns are encoded directly in the spatial structure of the device itself, the arrangement of beam splitters and phase shifters. A single pass through this static device unitarily transforms a reference input into the desired superposition of stored patterns, with no sequential $O(p)$ gates. When the number of stored patterns scales polynomially with $n$, the size of this spatial encoding remains efficient, and the preparation of the memory state is effectively $O(1)$ per use. Note that here and in the following, complexity refers to the number of sequential operations or repetitions required for one retrieval as a function of $n$, assuming that the pattern-dependent memory hardware has already been constructed. Thus a single traversal of a fixed photonic device is O(1) in this resource measure; its physical size and construction cost are outside the scope of the present analysis. While this architecture was proposed already in the 90s, the recent advances in integrated photonic circuits lead to devices with hundreds of stable, programmable modes \cite{harris}; further scaling is an engineering rather than a conceptual challenge. This single-pass property is essential to avoid re-preparation costs
and to realize a genuine associative memory. The analysis presented below, however, is independent of the specific physical implementation.

Quantum associative memories mirror the Hopfield paradigm perfectly. Both encode pattern in a Hamiltonian: in the classical Hopfield model retrieval is achieved through irreversible dynamics that relaxes the system toward one of the stored attractors. In the quantum memory, recall dynamics is unitary, followed by a measurement. 

Here we shall analyse quantitatively the trade-off between complexity and precision for the retrieval algorithm. We shall show that an exponential complexity advantage (square root speed-up when measured with respect to the total number of possible states) with respect to the classical brute force approach with simultaneous high precision of retrieval is possible for a number of patterns $p$ that scales polynomially in the number $n$ of qubits. A polynomial quantum advantage in capacity is achieved with respect to classical neural network implementations, with no cross-talk destroying the retrieval capabilities. The purpose of this paper is to derive analytically the algorithmic asymptotic scalings and trade-offs, leaving detailed implementation issues for future work. 

The first attempt to construct a quantum content-addressable memory \cite{ventura} was based on the Grover algorithm \cite{grover}, using a subset of the $n$ qubits as the memory address. This, however, is not yet an associative memory since it does not permit to retrieve noisy, corrupted inputs. The first associative retrieval algorithm was proposed in \cite{cat1, cat2, cat3}. The idea is to apply a unitary transformation that turns a uniform quantum superposition of the patterns into a pure thermal state in which the probabilities of the patterns in the superposition are governed by a Boltzmann distribution with an effective temperature given by the inverse number of ancilla qubits used and with ground state given by the queried input. With this construction, the complexity of retrieval does not depend on the number of patterns anymore, but only on the effective temperature. When the final state is measured, the probability of retrieval is highest for the patterns closest (in Hamming distance) to the input, the more so the lower the temperature. 

Let us start from a quantum memory state $|m \rangle $ encoding $p$ binary patterns $|p^i \rangle = |p^i_1 \dots p^i_n\rangle $ of $n$ qubits (note that the notation $p^i$ indicates the $i^{\rm th}$ element of the set of $p$ patterns),
\begin{equation}
|m\rangle = {1\over \sqrt{p}} \sum_{i=1}^p |p^i \rangle \ .
\label{memory}
\end{equation}
An efficient algorithm to load the patterns to form this state is described in \cite{cat1}; here we will not repeat the derivation but, rather take this state as our starting point. We just mention, for completeness, that the cost of this state-preparation step is $O(np)$ for $p$ patterns of $n$ bits each in the circuit model. For the associative retrieval algorithm we shall add a control register $|c\rangle$ with $b$ qubits all in state $|0\rangle$, so that the initial state becomes
\begin{equation}
|\psi_0\rangle = {1\over \sqrt{p}} \sum_{k=1}^p | i; p^k; 0_1,\dots,0_b \rangle \ ,
\label{init}
\end{equation}
where $|i\rangle = |i_1, \dots, i_n\rangle $ denotes the input qubits. 

Before describing retrieval in detail let us focus on the elementary quantum gates (for a review see \cite{nielsen}) used in this algorithm. First of all we will need the single-qubit NOT gate, represented by the first Pauli matrix $\sigma_1$ and the Hadamard gate H, with the matrix representation
\begin{equation}
H = {1\over \sqrt{2}} 
\begin{pmatrix} 
1 & 1\\
1 & -1 
\end{pmatrix} \ .
\label{adda}
\end{equation}
Moreover, we will use the two-qubit XOR (exclusive OR) gate, which performs a NOT operation on the second qubit if and only if the first one is in state $|1\rangle$. In matrix notation this gate is represented as ${\rm XOR} = {\rm diag} \left( I, \sigma_1 \right)$, where $I$ denotes a two-dimensional identity matrix. For all these gates we shall indicate by subscripts the qubits on which they are applied, the control qubits coming always first.  

Let us now focus on the retrieval algorithm itself. As a preliminary step, we apply the Hadamard gate to the first control qubit, 
\begin{eqnarray}
|\psi _1\rangle &&= {1\over \sqrt{2p}} \ \sum_{k=1}^p |i; 
p^k; 0_1,\dots ,0_b\rangle
\nonumber \\ 
&&+{1\over \sqrt{2p}} \ \sum_{k=1}^p |i; 
p^k; 1_1,\dots ,0_b\rangle \ .
\label{ad}
\end{eqnarray}
We now apply to this state the following $O(n)$ combination of quantum gates,
\begin{equation}
|\psi _2\rangle = \prod_{j=1}^n \ NOT_{m_j} 
\ XOR_{i_j m_j} |\psi _1\rangle \ .
\label{ae}
\end{equation}
As a result, the memory register qubits are in state $|1\rangle$ if $i_j$ and $p^k_j$ are identical
and $|0\rangle$ otherwise, 
\begin{eqnarray}
|\psi _2\rangle &&= {1\over \sqrt{2p}} \ \sum_{k=1}^p |i; 
d^k; 0_1,\dots ,0_b\rangle 
\nonumber \\
&&+{1\over \sqrt{2p}} \ \sum_{k=1}^p |i; 
d^k; 1_1,\dots ,0_b\rangle \ ,
\label{af}
\end{eqnarray}
where $d^k_j = 1$ if and only if  $i_j=p^k_j$ and $d^k_j=0$ otherwise.

We now introduce the Hamiltonian
\begin{eqnarray}
{\cal H} &&= \left( d_H \right)_m \otimes \left( \sigma_3 \right)_{c_1} \ ,
\nonumber \\
\left( d_H \right)_m && = \sum_{j=1}^n 
\left( {\sigma_3 + 1\over 2} \right) _{m_j}\ ,
\label{ag}
\end{eqnarray}
where $\sigma _3$ is the third Pauli matrix.
${\cal H}$ measures the number of 0's in the memory register $m$, with a plus sign if $c_1$
is in state $|0\rangle$ and a minus sign if $c_1$ is in state $|1\rangle$. This is the
number of qubits which are different in the input and memory registers $i$ and
$m$. This quantity is called the Hamming distance and it is the most common measure used for the similarity between binary patterns. 

The Hamiltonian (\ref{ag}) can be also be realized by simple gates involving only one or two qubits. To see this, let us introduce the single-qubit gate
\begin{equation}
U = 
\begin{pmatrix} 
{\rm e}^{i{2\pi\over n}}& 0\\
0 & 1 
\end{pmatrix} \ ,
\label{q}
\end{equation}
and the two-qubit controlled gate
\begin{equation}
CU^{-2} = |0\rangle \langle 0| \otimes 1 + |1\rangle \langle 1|
\otimes U^{-2} \ .
\label{r}
\end{equation}
It is then easy to check that ${\rm exp}\left( i\pi {\cal H}/2n \right)$ 
in eq. (\ref{af}) can be realized as a $O(n)$ product of simple gates as 
\begin{equation}
{\rm e}^{i{\pi \over 2n} {\cal H}} \ |\psi _2\rangle = 
\prod_{i=1}^n \left( CU^{-2} \right) _{c m_i} \ \prod_{j=1}^n U_{m_j}
\ |\psi_2\rangle \ ,
\label{t}
\end{equation}
where $c$ denotes the control qubit under consideration. This
means that one first implements globally ${\rm exp}\left( i\pi  d_H /2n \right)$
and then one corrects by implementing ${\rm exp}\left( -i\pi  d_H /n \right)$ on that part of
the quantum state for which the control qubit $|c\rangle$ is in state $|1\rangle$. 

Every term in the superposition (\ref{af}) is an eigenstate of ${\cal H}$ with a
different eigenvalue. Applying the unitary operator ${\rm exp} 
(i \pi {\cal H}/2n)$ to $|\psi _2\rangle$ one obtains
\begin{eqnarray}
|\psi _3\rangle &&= {\rm e}^{i{\pi \over 2n}{\cal H}} \ |\psi_2\rangle \ ,
\label{ah} \\
|\psi_3\rangle &&= {1\over \sqrt{2p}} \sum_{k=1}^p {\rm e}^{i{\pi\over 2n}
d_H\left( i, p^k\right)}
|i; d^k; 0_1,\dots ,0_b\rangle 
\nonumber \\
&&+ {1\over \sqrt{2p}} \sum_{k=1}^p {\rm e}^{-i{\pi\over 2n}
d_H\left( i, p^k\right)}
|i; d^k; 1_1,\dots ,0_b\rangle \ ,
\nonumber
\end{eqnarray}
where $d_H\left( i, p^k \right)$ denotes the Hamming distance bewteen the input
$i$ and the stored pattern $p^k$. 

Finally, we will restore the memory register to the initial state $|m\rangle$ by applying
the inverse transformation to eq. (\ref{ae}) and the Hadamard gate  
to the control qubit $c_1$. This costs further $O(n)$ gates and gives
\begin{eqnarray}
|\psi _4\rangle &&= H_{c_1} \prod_{j=n}^1 XOR_{i_j m_j}  
\ NOT_{m_j} \ |\psi_3\rangle \ ,
\label{ai} \\
|\psi_4\rangle &&= {1\over \sqrt{p}} \sum_{k=1}^p {\rm cos}\  {\pi \over
2n} d_H\left( i, p^k\right) 
|i; p^k; 0_1,\dots ,0_b\rangle 
\nonumber \\
&&+ {1\over \sqrt{p}} \sum_{k=1}^p {\rm sin}\  {\pi \over 2n}
d_H\left( i, p^k\right) 
|i; p^k; 1_1,\dots ,0_b\rangle .
\nonumber
\end{eqnarray}

At this point we repeat the exact same procedure sequentially for all $b$ control qubits $c_1$ to $c_b$. This gives 
\begin{eqnarray}
|\psi_{\rm fin}\rangle &&= {1\over \sqrt{p}} \sum_{k=1}^p \sum_{l=0}^b
\ {\rm cos}^{b-l} \left( {\pi\over 2n} d_H\left( i, p^k \right)\right) \times 
\nonumber \\
&&{\rm sin}^l \left( {\pi\over 2n} d_H\left( i, p^k \right)\right) 
\ \sum_{\left\{ J^l \right\}} |i; p^k; J^l\rangle ,
\label{al}
\end{eqnarray}
where $\left\{ J^l \right\}$ denotes the set of all binary numbers of $b$ bits with exactly $l$ bits 1 and $(b-l)$ bits 0. The total number of gates applied to
obtain this state scales as $O(nb)$. Given that the query input is fixed and the memory register is restricted to the fixed patterns, this state has support on $p 2^b$ basis states. The role of the 
$b$ control qubits is not to reduce the dimensionality of the Hilbert space, which remains of the same size, but to sharpen the amplitude distribution over it.

The final step of the retrieval algorithm uses the standard amplitude amplification procedure \cite{brassard}, where the reflection is performed with respect to the prepared state rather than the uniform superposition. Accordingly, the amplification acts on the full Hilbert space of memory and control registers, with the $l=0$ “good” subspace defined by all control qubits being in the zero state $|0_1 \dots 0_b\rangle$. The required reflections are implemented by conjugating a simple phase flip on a reference state by the unitary transformation already used in the algorithm. This implies 
\begin{equation}
{1\over \sqrt{P_{\rm good}}} = \sqrt{p\over \sum_k {\rm cos}^{2b} \left( {\pi \over 2n} d_H \left(i, p^k \right) \right)} 
\label{prob}
\end{equation} 
repetitions of an $O(b)$ oracle call. We shall assume, for simplicity of presentation and without restriction of generality, that the input $i$ consists of all zero qubits, $|i_j\rangle = |0\rangle$, $\forall j$. As a consequence, the Hamming distance $d_H \left(i, p^k \right)$ becomes simply the number of 1's in pattern $p^k$, which we will denote concisely by $d_{\rm H}^k$. 

From now we shall discuss the ``typical" case by assuming that all pattern bits are independently, identically distributed (``i.i.d."). This implies that the Hamming distance has a binomial distribution
\begin{equation}
d_{\rm H} \sim {\rm Bin} \left( n, {1\over 2} \right) \ ,
\label{bin}
\end {equation}
characterized by mean and standard deviation
\begin{equation}
\mu = {n\over 2} \ , \qquad 
\sigma = {\sqrt{n}\over 2} \ .
\label{meva}
\end{equation}

The complexity of the typical case is given by the expectation value of (\ref{prob}) obtained from the law of large numbers. Using the mean (\ref{meva}) as argument of the cos function in (\ref{prob}) gives a factor $2^{b/2}$. Recalling that $O(b)$ repetitions of the oracle call on an $n$-qubit state are required we get the overall complexity
\begin{equation}
{\cal C}_{\rm ret}= O\left( nb\right) 2^{b\over 2}  \ .
\label{lln}
\end{equation}
This is our first result: for a large number $p$ of patterns, the typical complexity of retrieval becomes independent of $p$, it is determined solely by the number $b$ of auxiliary qubits. We stress that this does not imply a speed-up beyond Grover’s bound for unstructured search. The advantage arises because associative retrieval is not an unstructured search problem: the distance-dependent weighting of patterns modifies entirely the structure of the problem, so that the complexity depends on the effective temperature parameter $b$ rather than directly on the number of stored patterns.

After amplitude amplification, the state is rotated onto the $l=0$ ``good" subspace. Measuring the control register and obtaining $|0_1 \dots 0_b \rangle$ leaves the memory register in a superposition of the stored patterns with amplitudes proportional to ${\rm cos}^b \left(\pi d_{\rm H}( i, p^k)/2n\right)$. Squaring and normalizing these amplitudes give the probability distribution
\begin{eqnarray}
P_b^k = P_b\left( p^k\right) &&= {1\over Z} \ {\rm cos}^{2b} \left( {\pi \over 2n} d_H^k \right) \ ,
\nonumber \\
Z &&= \sum_{k=1}^p {\rm cos}^{2b} \left( {\pi \over 2n} d_H^k \right) \ .
\label{ao}
\end{eqnarray}
This probability is peaked on those patterns which have the smallest Hamming distance to the input. The highest probability of retrieval by measurement of this 
rotated state is thus realized for the pattern which is most similar to the input. Here we stress our second result: there are never spurious memories created by cross-talk, like in 
the classical Hopfield model (for a review see \cite{muller}), and this is independent of the number $p$ of patterns.  

The role of the parameter $b$ becomes familiar upon realizing that the quantum distribution (\ref{ao}) is equivalent to a canonical Boltzmann distribution 
with (dimensionless) ``temperature" $t = 1/b$ and (dimensionless) energy levels 
\begin{equation}
E^k = -2 \ {\rm log} \ {\rm cos} \left( {\pi \over 2n} d_H^k \right) \ ,
\label{aq}
\end{equation}
with $Z$ playing the role of the partition function. The complexity of retrieval, thus is set only by the chosen temperature, which maximizes the retrieval probability for patterns with minimal Hamming distance to the input. In the following we derive how the precision of retrieval depends on this temperature. 

To do so we shall assume, for simplicity of presentation, that there is a ``ground-state pattern" identical to the input and we shall derive the influence of nearby ``distractor" patterns. This requires estimating the minimum distance $\delta^*n$ to the ground state pattern, defined by
\begin{equation}
p \ {\rm Prob} \left( d_{\rm H} \le \delta^* n \right) \simeq 1\ .
\label{mdd}
\end{equation}

We begin by the most desirable case of an exponential number of patterns, by setting $p=2^{\alpha n}$ with $\alpha < 1$. We now use Chernoff's bound for a binomial distribution (for a recent derivation see \cite{mulzer}) 
\begin{equation}
{\rm Prob} \left( d_{\rm H} \le \delta n \right) \le 2^{-\left[ 1-H\left(\delta\right) \right] n} \ ,
\label{cher}
\end{equation}
with 
\begin{equation}
H\left( \delta \right) =-\delta \  {\rm log} (\delta) -(1-\delta)\  {\rm log} (1-\delta) \ ,
\label{ent}
\end{equation}
the binary entropy associated with $\delta < 1$, to obtain the estimate that the typical minimal distance scales linearly with $n$, $\bar d_H = \delta^* n$ with
\begin{equation}
H\left( \delta^* \right) = 1-\alpha \ .
\label{est}
\end{equation}

The ratio of the retrieval probability of such a distractor to the ground-state probability is 
\begin{equation}
{\rm cos}^{2b} \left( {\pi \over 2} \delta^* (\alpha) \right) \ ,
\label{onedis}
\end{equation}
where $\delta^*(\alpha)$ is determined by the solution of (\ref{est}). The precision of retrieval is high when the total distractor probability (estimated as $p$ times (\ref{onedis})) is small
\begin{equation}
p \ {\rm cos}^{2b} \left( {\pi \over 2} \delta^* (\alpha) \right)  \ll 1\ . 
\label{manydis}
\end{equation}
This requires
\begin{equation}
b\gg {\alpha n \over -2 \ {\rm log} \ {\rm cos} \left( {\pi \over 2} \delta^* (\alpha) \right) } \ ,
\label{condexp}
\end{equation}
which shows that $b$ must scale linearly with $n$, $b=\beta n$, to guarantee a precise retrieval for exponential patterns. In this case, the complexity of retrieval (\ref{lln}) becomes
\begin{equation}
{\cal C}_{\rm ret} \simeq  O(n^2) \ 2^{{\beta \over 2}n}  \ ,
\label{expr}
\end{equation}
This is our third result: for an exponential number of stored patterns precision and efficiency of recall are incompatible. 

Let us now focus on a polynomial number of patterns, by setting $p=n^k$. To proceed we recall that the central limit theorem implies that
\begin{equation}
{\rm Bin } \left( n,{1\over 2} \right) \xrightarrow[\text{$n \gg 1$}]  \ {\cal N} \left( {n\over 2}, {\sqrt{n}\over 2} \right) \ .
\label{clt}
\end{equation}
Therefore, 
\begin{equation}
{\rm Prob} \left( d_H \le {n\over 2} - t \right) \simeq {\rm e}^{- {2t^2\over n}} \ ,
\label{pr}
\end{equation}
and the minimal distance $\bar d_H =n/2 -t^*$ is defined, in this case, by the equation 
\begin{equation}
n^k \ {\rm e}^{- {2{t^*}^2\over n}} = 1 \ ,
\label{mdp}
\end{equation} 
with solution 
\begin{equation}
{t^*} = O \left(\sqrt{n\  {\rm log} (n)} \right) \ ,
\label{sol}
\end{equation}
which is sublinear in $n$ and can, thus be neglected with respect to $n/2$, giving $\bar d_H \simeq n/2$. The total relative probability of distractors, in this polynmomial case, becomes
\begin{equation}
p \ {\rm cos}^{2b} \left( {\pi \over 4} \right)  = p \ 2^{-b} = {n^k \over 2^b} \ .
\label{dispol}
\end{equation}
Precision of retrieval is guaranteed if 
\begin{equation} 
b \gg k\ {\rm log} (n) \ .
\label{prpol}
\end{equation}
In this case, a number $b$ of auxiliary qubits scaling logarithmically with $n$ is sufficient for precise retrieval. Comparing with the complexity (\ref{lln}) we obtain our fourth result:  the goal of simultaneous high precision of retrieval and a quantum complexity advantage with respect to the classical brute force complexity $p$ can be achieved for 
\begin{eqnarray}
&&b = B \ {\rm log} (n) \ ,
\nonumber \\
&&{B\over 2} < k < B \ .
\label{efpr}
\end{eqnarray}
A polynomial capacity advantage with respect to neural network implementations is assured by the absence of cross-talk in the quantum associative memory. 

The precision analysis above concerns the benchmark case in which
one stored pattern is identical to the query, so that the Hamming
distance of the target pattern from the query is $r=0$. Thus, the
high-precision results derived in Eqs.~(20)--(34) are established
for this case. For such an exact query, the target pattern has Hamming distance zero
and therefore unit unnormalized probability. More generally, let
the target pattern be at Hamming distance $r$ from a corrupted
query. Its unnormalized probability is
\begin{equation}
P_{\rm target}\propto
\cos^{2b}\left({\pi r\over 2n}\right) \ .
\label{newa}
\end{equation}
Since the closest distractors for a polynomial number of random
patterns still have $\bar d_H\simeq n/2$, their total probability
relative to the target becomes
\begin{equation}
{P_{\rm distractors}\over P_{\rm target}}
\simeq
p\left[
{\cos(\pi/4)\over
 \cos\left(\pi r/2n\right)}
\right]^{2b}
=
{n^k\over
 2^b\cos^{2b}\left(\pi r/2n\right)}.
\label{newb}
\end{equation}
Writing $r=\rho n$, the precision condition can be expressed as
\begin{equation}
{n^k\over
 \left[2\cos^2\left(\pi\rho/2\right)\right]^b}
\ll 1 \ .
\label{newc}
\end{equation}
For every corruption rate $\rho<1/2$, one has
\begin{equation}
2\cos^2\left({\pi\rho\over2}\right)>1\ ,
\label{newd}
\end{equation}
so that a logarithmic number $b=O(\log n)$ of control qubits is
still sufficient. Consequently, retrieval retains polynomial
complexity for every target pattern that is closer to the query
than a typical random pattern.

Let us finally stress that quantum error correction could be important for protecting the memory state and the retrieval circuit against physical noise. It would not, however, remove the intrinsic precision-complexity trade-off obtained here, since it is derived under the assumption of ideal noiseless operations. More general control register architectures, non-uniform controlled phases, or optimized polynomial filters could potentially improve the retrieval algorithm. Whether such modifications can change the asymptotic scaling, however, remains an open question for future work.

To conclude let us mention that dense associative memories, also called modern Hopfield networks \cite{ramsauer}, are often referred to as formally capable of polynomial or exponential capacity in the number $n$ of neurons. However, since it corresponds to an increase of the number of parameters of the network, this enhanced capacity must be assessed together with the required pattern-dependent additional resources. The present quantum associative memory contains $n$ memory qubits and can retrieve any polynomial number $p=n^k$ of arbitrary patterns without cross-talk. Its advantage is therefore a polynomially enhanced capacity relative to the standard pairwise Hopfield model; a comparison of total hardware resources with modern Hopfield networks must additionally account for the pattern-dependent state-preparation device.

\section{Author contribution}
C. A. T. conceived the study, did the computations and wrote the manuscript.

\section{Declarations}

\subsection{Data availability}
Data sharing not applicable to this article as no datasets were generated or analysed during the current study.

\subsection{Funding} 
This research received no external funding.

\subsection{Conflict of interests}
The author declares he has no conflict of interests.

\subsection{Clinical trial}
Not applicable.

\subsection{Consent to publish declaration}
Not applicable.

\subsection{Ethics and consent to participate declarations}
Not applicable.

\end{document}